\documentstyle[12pt]{article}
 \begin{document}
\medskip
\begin{center} \large
               {\bf DYNAMICAL SPIN II} \\
	\normalsize \bigskip
	
	\bigskip \bigskip

Peter G.O. Freund\footnote{freund@theory.uchicago.edu}\\ 
\em{Enrico Fermi Institute and Department of Physics,University of Chicago, Chicago, IL 60637}\\ 

 \bigskip \bigskip \bigskip
 
 {\bf{ABSTRACT}}\\
  \end{center}
 
The possibility of building all particles from spinless constituents is explored. Composite 
fermions are formed from bosonic carriers of electric and magnetic charge of a composite abelian gauge field.
Internal attributes are accounted for by dimensional reduction 
from a higher-dimensional space-time in which the abelian gauge field is replaced by a composite higher-rank 
antisymmetric tensor field. The problem of building magnetically neutral fermions is considered.

\bigskip \bigskip  \bigskip \bigskip  

{\em It is with great sadness that I dedicate this paper to the memory of my friend Wolfgang Kummer. 
Wolfgang and I met during our student days in Vienna, and as fate would have it, we both went to Geneva for our 
first post-doctoral appointment. There we collaborated on the paper "The phases of the proton's 
electromagnetic form factors in the time-like region", Nuovo Cimento {\bf 24}, 1160 (1962).

During some work Wolfgang and I did in those early computer-era 
days, we were asked to use the computing facilities sparingly. For numerical integration we therefore availed
ourselves of the services of that marvelous one-of-a-kind CERN employee, Mr. Klein. This Mr. Klein could 
perform complex arithmetic operations in his head. A Holocaust 
survivor, he had used this unusual ability to earn his living in a circus in the postwar years. 
There he was spotted and recruited by CERN.
Mr. Klein calculated
our integrals to such accuracy, that in the end, the computer's streamlined task was reduced to not much more than 
confirming his estimates.

I cannot resist mentioning here another bond between Wolfgang and me: since we both grew up in Viennas, he 
in the Austrian capital and I in the Romanian city of Timi\c{s}oara, the Habsburg Empire's former "Little Vienna," 
we both developed keen musical interests, we both sang, we both were baritones, and we both gave recitals 
with mezzo-sopranos. 
Unlike me though, 
Wolfgang had as his singing partner none other than that wonderful mezzo-soprano Helga Dernesch, 
who after singing with him was destined to
become a major star of the Wiener Staatsoper. 

To this memorial volume I decided to contribute a paper I wrote in 1981, which however 
may still be of some interest today.
It has been available on SPIRES (EFI-81/07-CHICAGO, Feb 1981) for over a quarter of a century, 
but due to all kinds of complications it has never been published before. I chose 
this paper, because I clearly recall a discussion with Wolfgang about the ideas contained in it. 
In fact, his interest in this paper prompted me to give him the preprint of the original 1981 
version reproduced below without any changes.}

\newpage 

1. INTRODUCTION
\bigskip

The search for a simple way of accounting for the observed particle spectrum and 
interactions has led to ever more remote constituent and subconstituent models \cite{S}.  
In order to account for the observed fermions it is usually assumed that some or all 
of the constituents are themselves fermions, and thus carry half-odd-integer spin.  
Here we wish to explore the opposite case where none of the constituents carry spin 
so that all angular momentum is of dynamical origin.  Spinless bosons can bind into 
bosonic states of integer angular momentum.  If amongst these components there 
are gauge bosons, then we have the possibility of nontrivial topological objects that 
carry magnetic charge. Together with electrically charged objects we then have the 
ingredients to build spinorial fermions \cite{TF, JHG}.  For such a picture to make even remote 
phenomenological sense, a considerable "attribute" (i.e., flavor, color, etc) 
proliferation at the level of the electrically and magnetically charged constituents 
seems to be required.  An elegant way to avoid such a proliferation is provided by 
higher-dimensional Kaluza-type theories \cite{WKTF}.  Yet the idea of building fermions from 
electric and magnetic charges relies heavily on a 4-dimensional space-time.   To 
extend this idea to higher dimensions we propose to replace the abelian vector gauge 
field of 4 dimensions by gauge fields of higher (totally antisymmetric) tensorial rank 
\cite{KR}.  The corresponding carriers of electric and magnetic charge are then not point 
particles but extended objects, as we shall see.  Since abelian structures are natural 
in this context, the nonabelian gauge fields of electroweak and strong interactions 
are to be viewed as composites. Both Bose and Fermi composites being possible, 
dynamical supersymmetry may also arise. 

\bigskip \bigskip 

2. MAGNETIC CHARGES IN HIGHER DIMENSIONAL SPACES
\bigskip

Consider a Minkowski space $M_d$ with one time- and $d-1$ space-dimensions. Define over $M_d$ a 
rank-$n$ antisymmetric tensor potential $A_{\mu _1....\mu _n} ~~ (n\leq d-1)$ or, equivalently, the $n$-form
$A = A_{\mu _1....\mu _n} dx^{\mu _1}\wedge dx^{\mu _2} \wedge...dx^{\mu _n}$.  The field 
strengths are the components of the n+1-form $F=dA$. It's dual $*F$ is a $d-n-1$ form. 
Introducing the "electric" current $n$-form $J$ and the "magnetic" current $d-n-2$-form $K$, 
the field equations are 
$$
d*F = *J, ~~    dF = *K. 
$$
The $n$-form $J$ can be restricted to "live" on a $(d_e+1)-$-dimensional submanifold of $M_d$, provided $d_e+1 \geq n$.  
We shall consider here the "minimal" case $d_e+1 = n$, and specifically that one of the 
dimensions of the submanifold is time-like (a proper-time) and $d_e$ are space-like.  At 
any proper-time the support of the electric charge is then $d_e$-dimensional.  Similarly, 
the support of magnetic charge has at least $d_m=d-n-3$ dimensions.  Notice that
$$
d_e+d_m=d-4,                             \eqno{(1)}
$$
so that 
both pointlike electric and magnetic charges are possible only in 4-dimensions.   In 
general $d_e\neq d_m$, but in every even dimension there exists an electric-magnetic-dual case in 
which $F$ and *$F$ are both $\frac{d}{2}$ forms, so that $n=\frac{d}{2}-1$ and $d_e=d_m=\frac{d-4}{2}$. 
It is this electric-magnetic-dual case that interests us here.

At this point we want to make precise what we mean by an electric or a 
magnetic field configuration and to find the counterparts of the Coulomb-electric and 
Dirac-magnetic (monopole) potentials.  To this effect we first consider a static 
configuration such that at all times the support of $J$ is the $(\frac{d}{2}-1)$-hyperplane 
(our results are obviously generalizable to other J-supports)
$$
x^1=x^2=....=x^{\frac{d}{2}+1}=0.        \eqno{(2)}
$$

Here it is worthwhile to streamline our notation. The last coordinate $x^d$ is 
designated as time, the metric signature is thus (--...-+). Indices that range from $1$ to $\frac{d}{2}+1$ 
(from $\frac{d}{2}+2$ to $d$)
will be designated by letters from the beginning (middle) of the latin alphabet $a, b, 
c,....(m, n, p, ....)$.  Thus, e.g., the hyperplane equation (2) becomes $x^a=0$.  A set of  totally 
antisymmetrized indices of either type will be indicated in a generic way by a square 
bracket containing one of them.   Specifically, $[a]$ means $a_1 a_2...a_{\frac{d}{2}-1}$
with all a's ranging from $1$ to $\frac{d}{2}+1$,
and $[m]$ means $m_1 m_2 ... m_{\frac{d}{2}-1}$ with all m's ranging from $\frac{d}{2}+2$ to $d$.  
Finally, the Levi-Civit\`{a} symbol for the first $\frac{d}{2}+1$ (last $\frac{d}{2}-1$) indices will be written as
$\epsilon_{[a]bc}$ $(\epsilon _{[m]})$.

With this notation the only nonvanishing components of $J$ in our static 
situation (2) are given by 
$$
J_{[m]}=\epsilon _{[m]}\frac{e}{\Omega _{\frac{d}{2}}} \delta(x^1)... \delta(x^{\frac{d}{2}+1})   \eqno{(3a)} 
$$
where $\Omega _{\frac{d}{2}}$ is the $\frac{d}{2}$-dimensional total solid angle (area of 
unit $\frac{d}{2}$-sphere: $\Omega _2=4\pi ....$).  The field equations then yield 
$$
A_{[m]}=\frac{e}{r^{\frac{d}{2}-1}} (\frac{-2}{d-2}),    \eqno{(3b)}
$$
with 
$$
r^2=(x^1)^2+ (x^1)^2+...(x^{\frac{d}{2}+1})^2    \eqno{(3c)} 
$$

The nonvanishing field components are all "electric" and of the form 
$$
E_a=F_{a[m]}= \frac{e x^a \epsilon_{[m]}}{r^{\frac{d}{2}+1}}   \eqno{(3d)}
$$
independent of time and of the last $\frac{d-4}{2}$
space coordinates, as expected.   The equations (3) define a Coulomb-electric field 
configuration. A Dirac-magnetic configuration with support in the same hyperplane requires a structure of $K$ of the 
same type as Eq. (3a) for $J$ but with the "electric charge" $e$ replaced by the 
"magnetic charge" $g$.
	For the magnetic field 
$$
H_a=\frac{1}{\frac{d}{2}!} \epsilon _{ab[c]} F_{b[c]}= 
\frac{1}{(\frac{d}{2}-1)!}\epsilon _{ab[c]} \partial _b A_{[c]} \eqno{(4)} 
$$
we require it to be of the same form as the Coulomb 
field (3d) but with $e \rightarrow g$: 
$$
\frac{1}{(\frac{d}{2}-1)!}\epsilon _{ab[c]} \partial _b A_{[c]}= \frac{g x_a }{r^{\frac{d}{2}+1}}.  \eqno{(5)} 
$$

We now have to solve these equations for $A_{[a]}$.
As in the familiar 4-dimensional Dirac case, the Bianchi identities force us to 
introduce a string of singularities starting in each point of the support of K. For 
convenience we point all these strings along, say, the $3$-direction.  The proper Ansatz 
for $A_{[a]}$ is then
$$
A_{[a]}= \epsilon _{[a]3b}x^b f(r,\xi), ~~~ \xi=\frac{x^3}{r}.   \eqno{(6a)} 
$$Inserting this Ansatz into Eq. (5) we find 
$$
f(r,\xi)= r^{-\frac{d}{2}}F(\xi)   \eqno{(6b)} 
$$
with $F(\xi)$ obeying the 
differential equation 
$$
F'(\xi)-\frac{d}{2}\frac{\xi}{1-\xi^2}F(\xi) + \frac{g}{1-\xi^2}=0.  \eqno{(7)}
$$
Since $|\xi| \equiv |\frac{x^3}{r}| \leq 1$, it is convenient to 
introduce the variable 
$$
\theta = Arc cos \xi  \eqno{(6c)} 
$$
and the function 
$$
G(\theta) \equiv F(\xi)    \eqno{(6d)} 
$$
The solution to Eq. (7) is then
$$
G(\theta)= g (sin\theta)^{-\frac{d}{2}} [\int^\theta (sin\psi)^\frac{d-2}{2} d\psi + \lambda]  \eqno{(6e)}
$$
with $\lambda$ an integration constant that 
goes with the indefinite integral.  The equations (6) determine the Dirac potentials.  
As an example for $d=4$ we obtain the familiar Dirac result with the string along the 
positive (negative) $3$-axis for $\lambda= -1$ ($\lambda=+1$).  
From the familiar recursion formula for the indefinite 
integral in (6e), $G(\theta)$ is periodic in $\theta$ for $d$ an integer multiple of 4. For $d=2~(mod~4)$ 
the indefinite integral in $G(\theta)$ contains also a linear term in $\theta$ which can be brought 
to the main determination $0 < \theta <\pi$ by readjusting the integration constant $\lambda$.

	At this point we have to consider some global problems.  As defined above, the 
support of both electric and magnetic charges for even $d > 4$ are infinite $\frac{d-4}{2}$-dimensional 
hyperplanes, which is undesirable. But if the higher dimensions are to be 
unobservable, then $d-4$ space-like dimensions must have compact topology (e.g. a 
torus).  But as we saw, $d_e+d_m=d-4$, so that an electric-magnetic charge pair can always fit into 
the "extra" compact space-like dimensions.

	From the electric and magnetic charges in $d$ dimensions we can construct 
spinorial fermions. One way to see that, is to replicate the Tamm-Fierz [2] 
arguments for our spread-out charges. Heuristically, upon dimensional reduction 
(i.e., compactification of the $d-4$ dimensions in which the charges are extended) the 
$\frac{d}{2}-1$-tensor field contains ordinary $4$-dimensional abelian gauge fields. Spinors can then 
be constructed from Bose electric and magnetic charges in the usual way [2,3]. But 
these $4$-dimensional spinors must originate in $d$-dimensional spinors; they cannot 
come from $d$-dimensional tensors by dimensional reduction.  Both $e$ and $g$ have 
dimension of $(d$-dimensional action)$^\frac{1}{2}$ so that the ensuing $d$-dimensional Dirac 
quantization is meaningful.

\bigskip \bigskip

3. COMPOSITE PICTURE
\bigskip

The way to use the arguments above to construct composite models is as follows.  
Suppose one starts with a $d$-dimensional space-time $d$ = even integer larger than $4$.  In this 
space there exist a set of scalar fields for which one can build a composite $({\frac{d}{2}-1})$-rank 
antisymmetric tensor field or, alternatively, this field can be "elementary".  There 
can further appear electric and magnetic extended objects $\epsilon$ and $\mu$ and the corresponding 
anti-objects $\bar{\epsilon}$ and $\bar{\mu}$.  
From $\epsilon\mu, ~ \epsilon\bar{\mu}, ~ \bar{\epsilon}\mu, ~ \bar{\epsilon}\bar{\mu}$ 
one can construct spinor composites, from $\epsilon\bar{\epsilon}, ~ \mu\bar{\mu}$ tensor composites.  
With suitable dynamics these composites may exhibit a "dynamical" 
supersymmetry.  If $d$ is large this may involve higher rank tensors.  A dimensional 
reduction is precipitated one way or another  \cite{WKTF} and in four dimensions we have a 
proliferation of composites since each spinor and tensor from $d$ dimensions branches 
into many counterparts in $4$ dimensions (just as in extended supergravities).  In $4$
dimensions the spectrum is very rich, the simplicity is restored in $d$ dimensions.  
This picture is, of course, very similar to extended supergravity except that the 
gauged supersymmetry in the original $d$ dimensions is viewed as dynamical, thus 
allowing higher rank tensors and spin-tensors, or higher spins in $4$ dimensions.  As 
it stands, this picture has a serious flaw: all fermions 
$\epsilon\mu, ~ \epsilon\bar{\mu}, ~ \bar{\epsilon}\mu, ~ \bar{\epsilon}\bar{\mu}$
contain one unit each of electric and magnetic charge.  All fermions 
of the theory must contain an odd number 
of these basic fermions and, as such, must carry odd, and therefore non-vanishing, 
electric {\em and} magnetic charges.  Even though we have not as yet specified the 
detailed nature of the abelian gauge field in $4$ dimensions whose sources these 
charges are, this is a serious difficulty. 
	
We want to sketch here one possible way out.  Consider (in four-dimensional 
space-time) a spherical shell of uniformly distributed electric charge.  Classically this 
tends to explode and the Casimir effect is known to have the wrong sign \cite{B} and thus 
does not stabilize the configuration.  It has been noted recently by Agostinho Ferreira, 
Zimerman and Ruggiero \cite{AZR} that in a distribution of both electric and 
magnetic charge along a spherical shell the Casimir effect is stabilizing.  
Specifically, they consider a spherical shell that is a perfect magnetic conductor at 
its polar caps, and a perfect electric conductor on the "ring" between these caps:  On 
the ring is uniformly distributed the electric charge while the two polar caps support 
uniform distributions of magnetic charge $\tilde{g}$ and $-\tilde{g}$ respectively, so that the whole system 
is magnetically neutral.  Here the Casimir effect is stabilizing.  We observe that for 
this system the angular momentum does not vanish as it would, were the magnetic 
charges at the two polar caps to have the same sign. By adjusting $e$ and $\tilde{g}$, we can fix the 
total angular momentum at $\frac{\hbar}{2}$, as would befit a spinor (as a model for the electron such 
a semiclassical argument requires much too large a size).  One may object that each 
polar cap contributes to the total angular momentum, which violates angular 
momentum quantization (or equivalently, the Dirac quantization).  This can be 
circumvented by postulating that such "polar caps" can never be isolated, but must 
always come in like- or opposite-charged pairs, as if they were doublets of a confining 
$SU(2)$ gauge theory.  This is similar to what would happen in discussing usual Dirac 
quantization were one guaranteed that all magnetically charged particles are 
composites made of an even number of very closely bound inseparable constituents 
of equal magnetic charge.  Obviously then, the Dirac quantization for "monopoles" 
would translate into a quantization for the constituents. 

The challenge is now to 
construct a detailed model that implements the ideas presented above.

	Following the completion of this work I received a Trieste preprint IC/80/180 
from J.C. Pati, A. Salam and J. Strathdee, in which similar ideas are explored in a 
rather different way. 

\bigskip

{\bf Note (August 13, 2008)}:
 
The results reported in section 2 of this paper have also been obtained
independently by R. Nepomechie \cite{N} and by C. Teitelboim \cite{T}. 

The paper by Pati, Salam and Strathdee mentioned in the last sentence of the text has since appeared, \cite{PSS}.

The 1981 work reported here was supported in part by the NSF: Grant No. PH-4-78-23669.
I wish to thank Beth An Nakatsuka for her help with the retyping of the 1981 preprint.\\

\end{document}